\documentclass[fleqn,10pt]{wlscirep}
\usepackage[utf8]{inputenc}
\usepackage[T1]{fontenc}
\usepackage{bm}
\usepackage{pdflscape}
\usepackage{enumitem}
\usepackage{ulem}
\usepackage[tikz]{bclogo}
\usepackage{lineno}
\usepackage{url}
\usepackage[super,square,compress]{natbib}

\usepackage{hyperref}

\hypersetup{linkcolor=orange,citecolor=orange,urlcolor=orange}

\definecolor{bgblue}{RGB}{245,243,253}

\title{Neutrinos from explosive transients at the dawn of  multi-messenger astronomy}

\author[1,*]{Irene Tamborra}
\affil[1]{Niels Bohr International Academy \& DARK, Niels Bohr Institute, University of Copenhagen, Blegdamsvej 17, 2100 Copenhagen, Denmark}

\affil[*]{e-mail: tamborra@nbi.ku.dk}

\begin{abstract}
With the advent of time-domain astronomy and the game-changing next generation of telescopes, we have unprecedented opportunities to explore the most energetic events  in our Universe through electromagnetic radiation, gravitational waves, and neutrinos. These  are  elementary particles, which exist in three different flavors and change the latter as they propagate in the dense core of astrophysical sources as well as en route to Earth.  To capitalize on existing and upcoming multi-messenger opportunities,  it is crucial to understand: 1.~the role of neutrinos in explosive transient sources  as well as in the synthesis of the elements heavier than iron;  2.~the impact of neutrino physics on the multi-messenger observables; 3.~the  information on the source physics carried by the detectable neutrino signal. In this review, the status of this exciting and fast-moving field  is outlined, focusing on astrophysical sources linked to collapsing massive stars and 
neutron-star mergers. In  light of   the upcoming plethora of multi-messenger data, outstanding open issues concerning the optimization of multi-messenger detection strategies are discussed.    
\end{abstract}
\begin{document}

\flushbottom
\maketitle

\thispagestyle{empty}

\section{Introduction}

We are used to  exploring the Universe through electromagnetic radiation. But now we can capture multiple cosmic messengers, including neutrinos, gravitational waves, and cosmic rays. Each messenger can provide complementary insight into the source physics. This multi-messenger approach is revolutionizing our understanding of the Universe.
Neutrinos play a key role in this context~\cite{Vitagliano:2019yzm}. They are elementary particles copiously produced in the compact cores and outskirts of dense astrophysical sources. Due to the weakness of their interactions, neutrinos are crucial messengers of the physics of  stellar interiors, since they can travel large distances undisturbed unlike photons. At the same time, the abundance of neutrinos in dense astrophysical sources makes their overall interaction rate large, with implications on the source physics and the nucleosynthesis taking place in the source itself. 

Rapid progress has occurred in neutrino astronomy in the past decade, with the discovery  of astrophysical neutrinos with energy above $\mathcal{O}(10)$~TeV from the IceCube Neutrino Observatory~\cite{IceCube:2016zyt}, as well as the growing number of  likely joint detections of electromagnetic and neutrino signals from blazars (e.g.~TXS 0506+056)~\cite{IceCube:2018dnn}, active galaxies (such as NGC 1068)~\cite{IceCube:2022der}, and our own Galaxy~\cite{IceCube:2023ame}. In addition, the first detection of gravitational waves from the merger of two neutron stars in 2017 was followed up by electromagnetic signals across wavebands coming from the associated gamma-ray burst  and the kilonova  (i.e., the bright electromagnetic transient  powered by the radioactive decay of   isotopes heavier than iron synthesized in the merger)~\cite{Margutti:2020xbo,Nicholl:2024ttg}. 
Neutrinos were not detected in coincidence with the gravitational-wave event GW 170817~\cite{ANTARES:2017bia,Super-Kamiokande:2018dbf}, in agreement with theoretical expectations. However, neutrinos are essential characters in the physics of neutron-star merger remnants, indirectly affecting the multi-messenger emission~\cite{Foucart:2024cjr}.
 The modeling of neutrino physics and weak interactions is swiftly advancing 
in  light of our renewed understanding of the impact that such physics has on the remnants of neutron-star mergers  as well as in the explosive death of  massive stars, and nucleosynthesis~\cite{Ehring:2023abs,Nagakura:2023mhr,Wu:2017drk,Just:2022flt,Li:2021vqj}.

The growing number of multi-messenger detections calls for an advanced modeling of the source and related particle production, as in most cases inconsistencies arise between  na\"ive  theoretical models and observations~\cite{2024Natur.626..742Y,Sneppen:2023vkk,Padovani:2024ibi}. In addition, the advent of time-domain astronomy has led to the discovery of  new classes of astrophysical transients with puzzling features, such as fast blue optical transients~\cite{Drout:2014dma,Ho:2021fyb}, which could be neutrino emitters~\cite{2018ATel11785....1B}; these sources are characterized by a lightcurve with a rise time faster than the typical supernova one, high peak luminosity, and their photon spectra appear blue. The multi-messenger signals arising from  the engulfment of a neutron star or a black hole  by a giant star will be useful to shed light on the mechanisms powering subsets of observed transients~\cite{Fryer:1998bh,Martinez-Mirave:2025pnz,DeMarchi:2021vwr,Metzger:2022xep,Soker:2022gna}. 
The understanding of the physics governing these  objects is crucial to grasp  how massive stars end their lives and whether these sources could act as additional reservoirs of chemical elements.

In  light of these exciting major theoretical developments and the  advent of large-scale observational facilities, this review outlines the state-of-the-art in neutrino astrophysics within a multi-messenger framework, it is arranged as follows. In Sec.~\ref{sec:SN},  the theory underlying the collapse of a massive star, the physics of neutron-star mergers, as well as recent developments on the role of neutrinos in  these sources are discussed. The features of the detectable thermal [$\mathcal{O}(10)$~MeV] neutrino signal are also explored, together with  what we could potentially learn from such  detections. 
Section~\ref{sec:outskirt} focuses on the non-thermal, high-energy [$10^3$--$10^9$~GeV] neutrino emission from sources stemming from collapsing massive stars. The  source regions where efficient particle acceleration could take place are highlighted, jointly with  how the detection of high-energy neutrinos could be instrumental to discriminate the  processes powering the source.  Multi-messenger detection strategies are outlined. Finally,  future challenges and opportunities are discussed in Sec.~\ref{sec:conclusions}.

\section{Neutrinos from the interiors of explosive transients}
In this section, an overview of the supernova mechanism is provided, followed by a discussion on neutrino flavor conversion in  core-collapse supernovae and its impact on the neutrino-driven explosion mechanism. Then, the signatures of the source physics imprinted in the  neutrino signal expected from a supernova burst as well as the ingredients entering the cumulative  neutrino flux from all stellar collapses in the Universe  are outlined.
Neutrinos are also produced in neutron-star mergers copiously. After summarizing the  physics of neutron-star mergers, we review the indirect impact of  neutrinos on the multi-messenger observables and  neutrino diffuse emission. 

\label{sec:SN}
\subsection{Core-collapse supernovae}

\subsubsection{The core-collapse supernova mechanism}
Stars with a mass larger than $8\ M_\odot$ evolve in hydrostatic equilibrium, developing an onion-shell structure, with increasingly heavier chemical elements synthetized through  nuclear burning.
Once iron nuclei are formed, 
the stellar core loses the nuclear energy support,   
and  energy to the stellar core can only be provided by stellar contraction. 
At first, the degenerate stellar core is stable against the gravitational collapse.
As the degenerate core approaches the Chandrasekhar limit, the gravitational collapse sets in. For collapsing stars with zero-age-main-sequence (ZAMS) mass $M_{\rm{ZAMS}} \lesssim 9 M_\odot$, the core is mainly made of oxygen, neon, and magnesium; the gravitational instability of the stellar core is set off by  electron capture by nuclei. For more massive  stars with iron cores, the instability is triggered by the  photo-disintegration of iron nuclei, due to the higher core temperature and pressure.
Once the collapse begins, the subsequent  electron captures (on nuclei, $(Z,A)+e^{-}\rightarrow (Z-1,A)+\nu_e$ with $Z$ and $A$ being the atomic and mass numbers,  and free protons) continue to neutronize  the core and reduce the electron degeneracy pressure, leading to a runaway collapse. 
During the core collapse, the inner part of the stellar core (where the baryonic density exceeds  $\gtrsim 10^{12}$~g cm$^{-3}$) contracts subsonically as a homologous core, while the outer part undergoes supersonic collapse. 
The infall stops when super-nuclear densities are reached in the inner 
core, which is mostly populated by nucleons.
At this point, a shock wave forms on the surface of the homologous core. 
The  shock  propagates  outwards into the supersonically infalling  matter. Meanwhile,   the homologous core and mass  falling through the shock lead to the formation of a neutron star, which  cools via neutrino emission.
As the shock propagates, it dissociates matter into free nucleons, losing 
 energy. The electron capture by free protons behind the shock is responsible for the emission of a burst of electron neutrinos: the neutronization burst. 
After a few tens of milliseconds since the core bounce, the shock stagnates as
a standing accretion shock.  
 To revive the shock, 
the delayed neutrino mechanism is considered to be the driver of the vast majority of supernova explosions. Only a small fraction of stellar collapses are  magneto-rotationally driven. In addition, magnetic fields can act as an additional energy reservoir in the gain region, aiding the neutrino-driven mechanism.

According to the neutrino-driven mechanism~\cite{Janka:2006fh,Burrows:2020qrp},  about one percent of the binding energy 
is deposited by neutrinos (the gravitational binding energy of the collapsing massive star is  $\sim 3 \times 10^{53}$~ergs, with  $99\%$ of this energy being carried by neutrinos)  in the region behind the stagnant shock 
through electron (anti)neutrino captures on free neutrons (protons) 
until  matter is heated up so much as to overcome the ram pressure of the infalling layers. 
Large scale asymmetries form in the postshock region due to hydrodynamic instabilities  (e.g., convective overturns of neutrino-heated gas and the standing accretion shock instability, SASI)~\cite{Bethe:1990mw,Blondin:2002sm}. Such instabilities  support the shock revival, pushing it outwards and enabling the runaway shock expansion responsible for the supernova explosion. 
Neutrinos also drive their  hydrodynamical instability during the stalled-shock phase: the lepton-number emission self-sustained asymmetry (LESA)~\cite{Tamborra:2014aua}. In the presence of  LESA, the flux of electron neutrinos minus antineutrinos develops a large-scale emission asymmetry, with implications  for  neutron-star kicks,  flavor conversion,  and the neutron-to-proton ratio in the neutrino-heated outflows.

The extreme speed observed for pulsars~\cite{Hobbs:2005yx} hints that neutron stars should have natal kicks due to asymmetric ejection of matter and neutrinos. Hydrodynamic simulations of the core collapse show that hydrodynamic and neutrino kicks (e.g.~due to LESA) contribute to the natal kick of the  compact remnant~\cite{Janka:2024xbp,Burrows:2023ffl}.   The recent electromagnetic detection of the inert (i.e.~X-ray  dormant) stellar-mass black hole binary VFTS 243  allows for  the first direct inference of the (neutrino) natal kick~\cite{Vigna-Gomez:2023euq}.

According to the magneto-rotational mechanism~\cite{Muller:2024slv}, if the proto-neutron star is fast rotating, its  rotational energy  can be tapped by magnetic fields.  The latter can be  amplified by convective dynamo,  turbulent flows, and  magnetorotational instabilities until the  energy required to revive the shock is achieved. In some cases, jets can be launched, driving highly asymmetric stellar explosions. If the jet launching is not successful, the jet may be choked in the dense stellar envelope, with no or dim electromagnetic signatures. 
Explosions driven by rotation and magnetic fields are expected to lead to superluminous supernovae and certain gamma-ray bursts~\cite{Mosta:2015ucs,Obergaulinger:2020cqq}. It remains to be understood up to which level assumptions on the pre-collapse magnetic field profile could affect the post-collapse dynamics of the  massive star.

\begin{figure}[ht]
\centering
\includegraphics[width=.455 \linewidth]{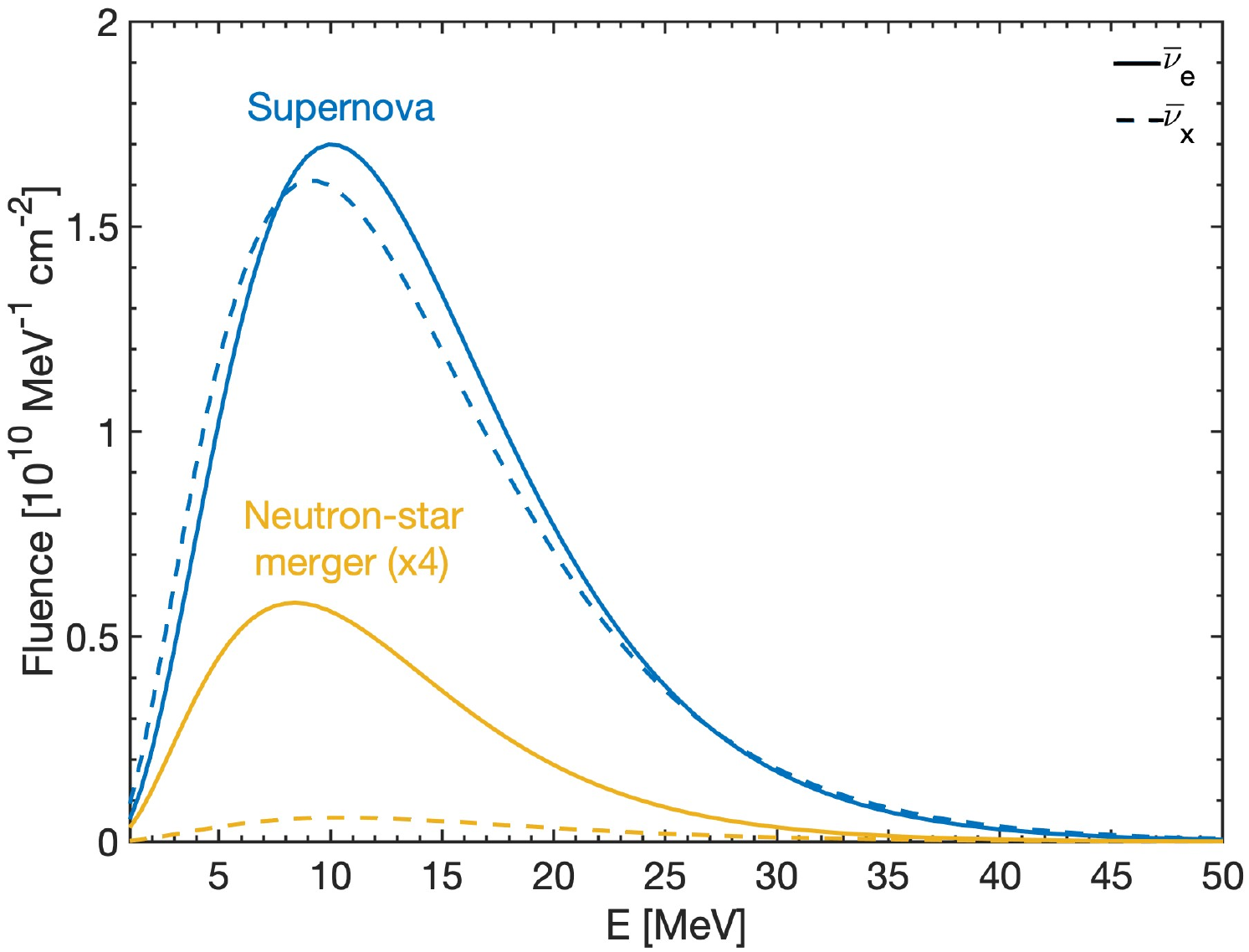}
\includegraphics[width=.46 \linewidth]{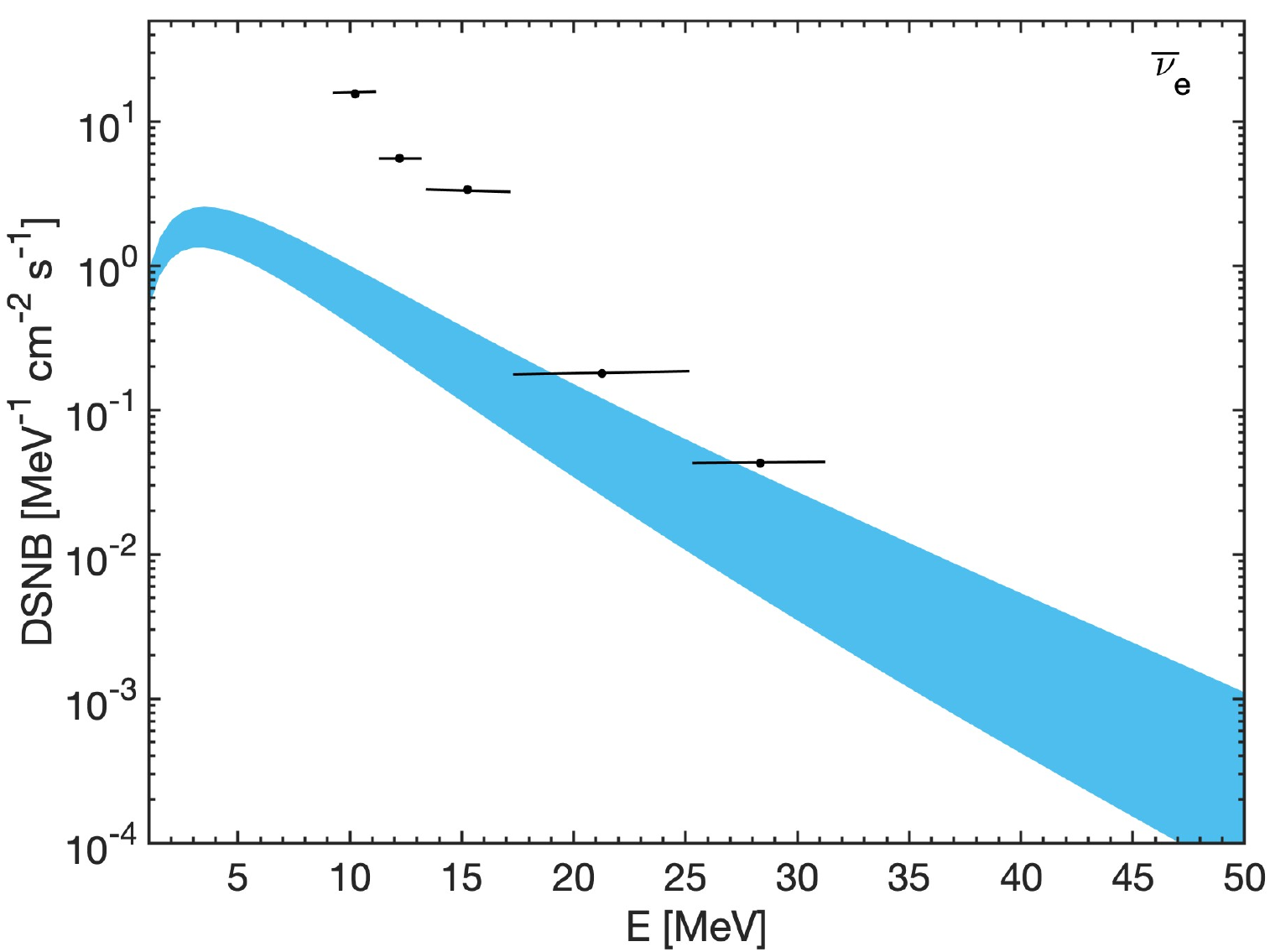}
\caption{Antineutrino fluence  from a supernova and a neutron-star merger (left panel)  and cumulative flux of neutrinos from all  supernovae in our Universe (right panel) . {\it Left}: Fluences of $\bar\nu_e$ (solid lines) and $\bar\nu_x$ ($=\bar\nu_\mu$ or $\bar\nu_\tau$, dashed lines) from a core-collapse supernova (in blue) and a neutron-star merger remnant (in orange, multiplied by  $4$ for visual purposes), both located at $10$~kpc from Earth. 
All curves have been obtained neglecting flavor conversion. The core-collapse supernova model is a $20~M_\odot$ progenitor~\cite{Garching} with SFHo nuclear equation of state~\cite{Steiner:2012rk}. The  merger remnant model is  an asymmetric merger of two neutron stars with masses of  $1.2$ and $1.5~M_\odot$ and SFHo nuclear equation of state~\cite{Kiuchi:2022nin}. For our models, the $\bar\nu_e$ fluence from neutron-star merger remnants  is about one order of magnitude smaller than the supernova one; however,  neutron-star mergers are not expected to occur as frequently as supernovae.   {\it Right}: Diffuse supernova neutrino background  in the electron antineutrino channel  as a function of the neutrino energy.
The band is representative and obtained allowing for  uncertanties on the supernova rate and fraction of black hole forming collapses~\cite{Moller:2018kpn}. The black error bars
represent the  upper limits from Super-Kamiokande loaded with Gadolinium~\cite{SK-Gd:2024}. The detection of the DSNB is expected in the near future, with the Super-Kamiokande   limits touching the theoretical predictions. }
\label{fluence_DSNB}
\end{figure}

During the stellar collapse, we expect neutrinos of all flavors to be emitted for $\mathcal{O}(10)$~s and with an average energy of $\mathcal{O}(10)$~MeV~\cite{Mirizzi:2015eza} (cf.~Table~\ref{tab:summary} for a summary of the multi-messenger emission from a core-collapse supernova; an overview of the neutrino  absorption, production, and direction-changing processes shaping the neutrino distributions in core-collapse supernovae is provided in~\cite{Burrows:2004vq,Lohs:2015bim}). Gravitational waves should be  emitted approximately in the same timeframe as MeV neutrinos.  The neutronization burst of electron neutrinos is expected to be visible for the first   tens of milliseconds. Then, during the accretion phase, electron neutrinos and antineutrinos are   emitted with roughly similar  luminosities, which are larger than the non-electron neutrino ones. As the newly born proto-neutron star cools and deleptonizes, after the explosion, (anti)neutrinos of all flavors are emitted with comparable luminosity. The antineutrino fluence  at Earth from a core-collapse supernova in our Galaxy  is shown in blue in the left panel of Fig.~\ref{fluence_DSNB}  as a function of the neutrino energy;  for simplicity,  flavor conversion (see Sec.~\ref{sec:QKE}) is neglected; the fluence at Earth should be a  combination of the non-oscillated ones represented in Fig.~\ref{fluence_DSNB}.

\begin{table}
 \caption{Multi-messenger transients stemming from supernovae (SN) and neutron star (NS) mergers, expected to be neutrino emitters, including gamma-ray bursts (GRB) as well as luminous fast blue optical transients (FBOT).  
 From left to right, the local rate is listed together with the  electromagnetic (EM) luminosity  with  the related wavebands  in parenthesis, the thermal (th) and non-thermal (non-th) neutrino luminosities in their characteristic energy range, as well as the  gravitational wave  (GW) energy in the associated frequency range.
}
    \centering
    \begin{tabular}{|c|c|c|c|c|c|c|}
    \hline
       Source  & Local rate  & EM luminosity  &  Th.~$\nu$  luminosity   & Non-th.~$\nu$  luminosity   &  GW energy  \\
         &  [Gpc$^{-3}$ yr$^{-1}$] & [erg s$^{-1}$] & [erg s$^{-1}$] &  [erg s$^{-1}$]   & [erg]  \\
       \hline 
       \hline
     Type  IIp SN    & $4.4\times 10^4$~\cite{Smith:2010vz}  & $10^{42}$~\cite{Zha:2023fmu}  & $10^{53}$~\cite{Mirizzi:2015eza} & $3.3 \times 10^{40}$~\cite{Sarmah:2022vra}  & $10^{44}$--$10^{48}$~\cite{Szczepanczyk:2021bka} \\ 
      & & [UVOIR, gamma-rays] & $[1$--$10^2$~MeV] & $[1$--$10^8$~GeV] &  [$10^2$--$10^3$~Hz] \\
   \cline{1-3}  \cline{5-5}
     Type IIn SN    & $8 \times 10^3$~\cite{Smith:2010vz}  & $10^{43}$~\cite{Salmaso:2024jry}  &   & $3.3 \times 10^{41}$~\cite{Sarmah:2022vra}  &   \\
       &   & [UVOIR, gamma-rays]  & & $[1$--$10^8$~GeV]  &   \\
     \cline{1-3}  \cline{5-5}
      Type Ib/c SN   & $3 \times 10^4$~\cite{Li:2010kc} & $10^{43}$~\cite{Sollerman:2021svd}  & & $6.7 \times 10^{33}$~\cite{Sarmah:2022vra}  &   \\
         &   &  [optical, gamma-rays] & & $[1$--$10^8$~GeV]  & \\
    \cline{1-3}  \cline{5-5}
       Superluminous     & $91$~\cite{Prajs:2016cjj}  & $10^{43}$--$10^{44}$~\cite{Gal-Yam:2018out}  &  & $10^{41}$~\cite{Pitik:2023vcg}  &   \\
       SN    &   &    [optical, radio/X-rays] & & $[10^2$--$10^7$~GeV]   &   \\
     \hline
        NS merger \&      &  $10$--$1700$~\cite{KAGRA:2021duu} & $10^{41}$--$10^{42}$~\cite{Drout:2017ijr}        & $1.5 \times 10^{53}$~\cite{Wu:2017drk} & $5\times 10^{47}$~\cite{Kimura:2018vvz}  & $\lesssim 2.2\times 10^{53}$~\cite{Zappa:2017xba} \\
        kilonova     &  $\lesssim 900$~\cite{Andreoni:2021ykx} & UVOIR      & $[1$--$10^2$~MeV] &  $[10^2$--$10^7$~GeV] & $[10^3$~Hz]~\cite{LIGOScientific:2017zic}  \\
        \hline
            Long GRB   &  $1.5$~\cite{Lan:2019nsa} & $10^{51}$~\cite{Pescalli:2015yva}    & N/A & $6 \times 10^{50}$~\cite{Kimura:2022zyg}  & $10^{50}$~\cite{Gottlieb:2022qow}    \\
       &  &     [X-rays, gamma (prompt)  & & $[10^2$--$10^8$~GeV] & $[10$--$10^2$~Hz]    \\
        &  & 
        broadband (afterglow)]   &   &  & \\
         \cline{1-3} \cline{5-5}
        Short GRB     &  $150$~\cite{Zevin:2022dbo} & $10^{52}$~\cite{Ghirlanda:2016ijf}   &   & 
        $10^{42}$~\cite{Biehl:2017qen}   &   \\
        &  &      [X-rays, gamma (prompt)  &   & $[10^2$--$10^8$~GeV] &    \\
         &  &    broadband (afterglow)]  &   & &   \\
         \hline
     Luminous        & 300~\cite{Ho:2021fyb}   & $10^{44}$~\cite{Ho:2021fyb,Coppejans:2020nxp}  & N/A & $\lesssim 10^{51}$~\cite{Stein:2019ivm}   & N/A \\
      FBOT      &      & [UVOIR, X-rays, radio]  & & $[10^2$--$10^7$~GeV]  & \\
     \hline
    \end{tabular}
    \label{tab:summary}
\end{table}

The SN 1987A remains the only core-collapse supernova detected in neutrinos up to date~\cite{Koshiba:1992yb,Kamiokande-II:1987idp,Bionta:1987qt}. While only approximately two dozen neutrino events were gathered in 1987, 
these data have been crucial showing broad agreement  with theoretical expectations;  the observed duration of the neutrino burst provided the first evidence of neutrino diffusion out of a dense core.
Recent work has given a fresh look into our understanding of the collected neutrino events, confirming that the theoretical models agree well with the  SN 1987A neutrino signal detected during  the first seconds~\cite{Fiorillo:2023frv}. 
However, the  neutrino events observed around $10$~s are still challenging to explain, since state-of-the-art supernova simulations including proto-neutron-star convection and nucleon correlations in the opacities of neutrinos predict a shorter cooling phase.

\newpage
  \subsubsection{Flavor conversion of neutrinos}
  \label{sec:QKE}
 \begin{bclogo}[
  couleur=bgblue,
  arrondi=0,
   logo= ,
  barre=none,
  noborder=true]{Neutrino quantum kinetics}
  \indent  When the density of neutrinos is large, as it is the case  in the proximity of the proto-neutron star,  neutrinos interact among themselves, in addition to interactions with the matter background.  
  To model this phenomenon, for a time $t$, location $\mathbf{x}$, and momentum $\mathbf{p}$, the neutrino mean field is represented through  a $3\times 3$ density matrix in the flavor basis [${\mathcal{\rho}}(t, \mathbf{x}, \mathbf{p})$ for neutrinos and ${\bar{\varrho}}(t, \mathbf{x}, \mathbf{p})$ for antineutrinos]. The diagonal elements of the density matrix represent the occupation numbers linked to the phase-space densities and the off-diagonal elements track  flavor coherence. The equations of motion for neutrinos and antineutrinos are, respectively~\cite{Sigl:1993ctk}:
  \begin{eqnarray}
 \centering
      &[\partial_t + \mathbf{v}\cdot \nabla_\mathbf{x} + \mathbf{F}\cdot \nabla_\mathbf{p}] \varrho_{\mathbf{x},\mathbf{p},t}& =  - \imath [\mathcal{H}_{\mathbf{x},\mathbf{p}, t}, \varrho_{\mathbf{x},\mathbf{p},t}] + \mathcal{C}[\varrho_{\mathbf{x},\mathbf{p},t}, \bar{\varrho}_{\mathbf{x},\mathbf{p},t}]\ ,\\    
      &[\partial_t + \mathbf{v}\cdot \nabla_\mathbf{x} + \mathbf{F}\cdot \nabla_\mathbf{p}]\bar{\varrho}_{\mathbf{x},\mathbf{p}, t}& =  -\imath [\bar{\mathcal{H}}_{\mathbf{p},\mathbf{x}, t}, \bar{\varrho}_{\mathbf{x},\mathbf{p}, t}] + \bar{\mathcal{C}}[\varrho_{\mathbf{x},\mathbf{p},t}, \bar{\varrho}_{\mathbf{x},\mathbf{p}, t}]\ .
  \end{eqnarray}
For each equation, the terms on the left hand side  take into account neutrino propagation (with $\mathbf{v} = \mathbf{p}/|\mathbf{p}|$ being the unit vector representing the direction of the velocity of the neutrino field) and the impact of an external force ($\mathbf{F}$, such as the gravitational field) on neutrino propagation;  on the right hand side, the functionals  $\mathcal{C}$ and $\bar{\mathcal{C}}$  encapsulate details on non-forward collisions of neutrinos with other neutrinos as well as with  the medium (i.e., absorption, emission, and direction-changing processes).  

The Hamiltonian matrix in the commutator is
 $\mathcal{H}_{\mathbf{x},\mathbf{p}, t} = \mathcal{H}_{\mathrm{vac}} + \mathcal{H}_{\mathrm{mat}} + \mathcal{H}_{\nu\nu}$.
The vacuum term of the Hamiltonian ($\mathcal{H}_{\mathrm{vac}}= \mathcal{M}^2/2E$) depends on  $\mathcal{M}^2$ which is the  matrix of squared neutrino masses  responsible for vacuum oscillations and has opposite sign for neutrinos and antineutrinos. The matter term [$\mathcal{H}_{\mathrm{mat}} = \mathrm{diag}(\sqrt{2} G_F n_e, 0)$] is a function of the  effective electron density $n_e$. The {\it Mikheyev, Smirnov and Wolfenstein (MSW) resonant flavor conversion} of (anti)neutrinos in matter occurs when  the diagonal components of the matter and vacuum Hamiltonians become comparable. (Note that, in this basis, $\mathcal{H}_{\mathrm{vac}}$ and $\mathcal{H}_{\mathrm{mat}}$  have the same  sign for neutrinos and opposite one for antineutrinos~\cite{Duan:2006an}.)  The {\it neutrino self-interaction} term, $\mathcal{H}_{\nu\nu} = \sqrt{2} G_F (n_{\nu}+n_{\bar\nu}) \int d\mathbf{p}^\prime (1-\mathbf{v}\cdot\mathbf{v}^\prime) [{\varrho}_{\mathbf{p},\mathbf{x}, t}-\bar{{\varrho}}_{\mathbf{p},\mathbf{x}, t}]$, depends on the (anti)neutrino number density for all flavors $n_{\nu, (\bar\nu)}$; the current-current nature of low-energy weak interactions is taken into account  through the term $(1-\mathbf{v}\cdot\mathbf{v}^\prime)$.
The non-linear flavor evolution induced by neutrino-neutrino interactions is crucially affected by the neutrino angular distributions, in addition to their energy distributions. 
\end{bclogo}

Due to the large neutrino density  in the proximity of the neutrino decoupling region, the coherent forward scattering of neutrinos on each other 
can trigger flavor conversion, if 
a crossing in the angular distributions of the electron neutrino lepton number occurs.
The conversion rate  depends on  the (anti)neutrino number density and has been dubbed ``fast,''  since this flavor conversion can potentially exist in the limit of vanishing neutrino masses~\cite{Izaguirre:2016gsx,Chakraborty:2016yeg}; this is different from the ``slow'' neutrino-neutrino flavor conversion that  requires neutrino mass splittings.

\begin{figure}[t]
\centering
\includegraphics[width=.45 \linewidth]{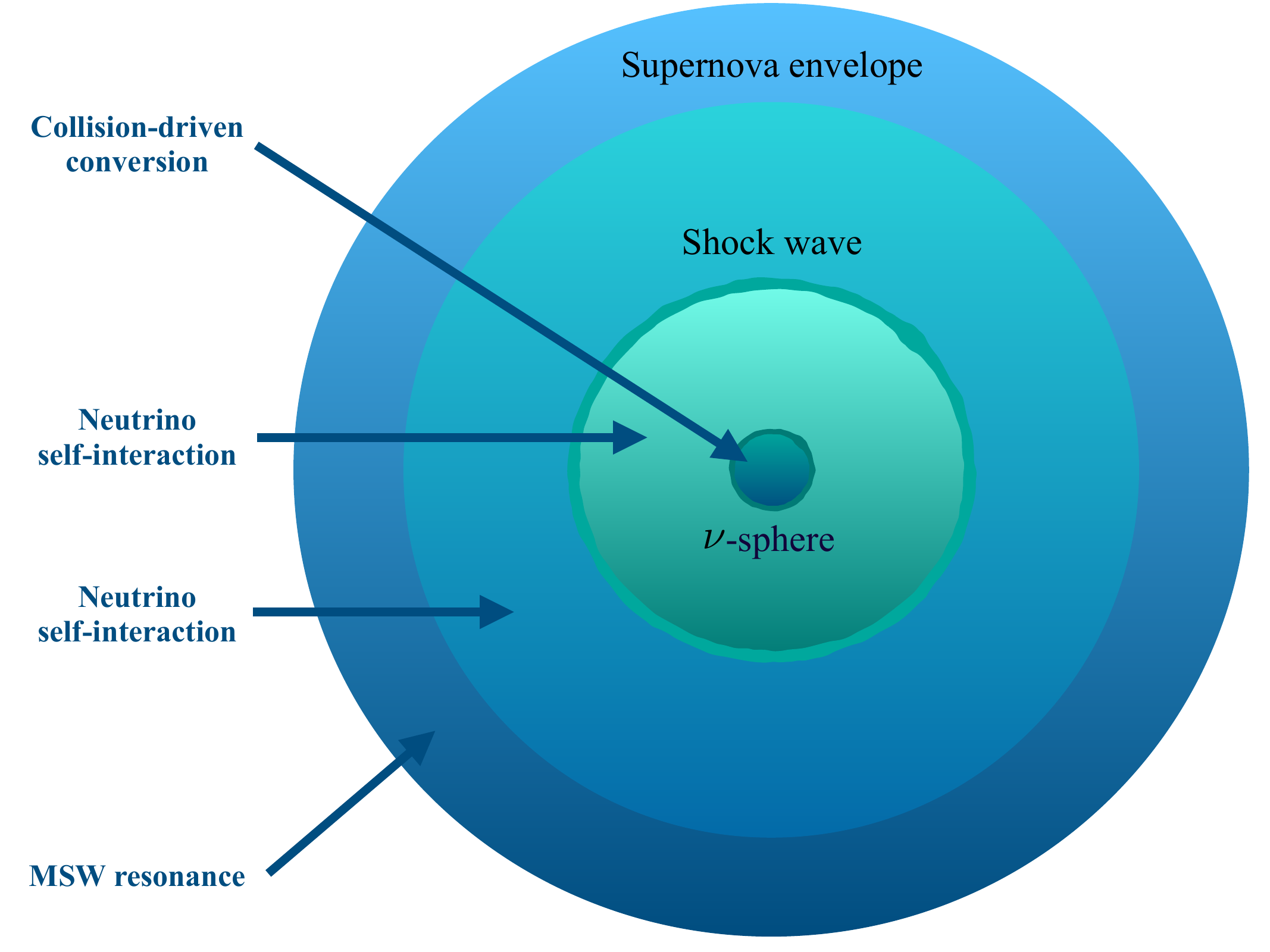}
\includegraphics[width=.45 \linewidth]{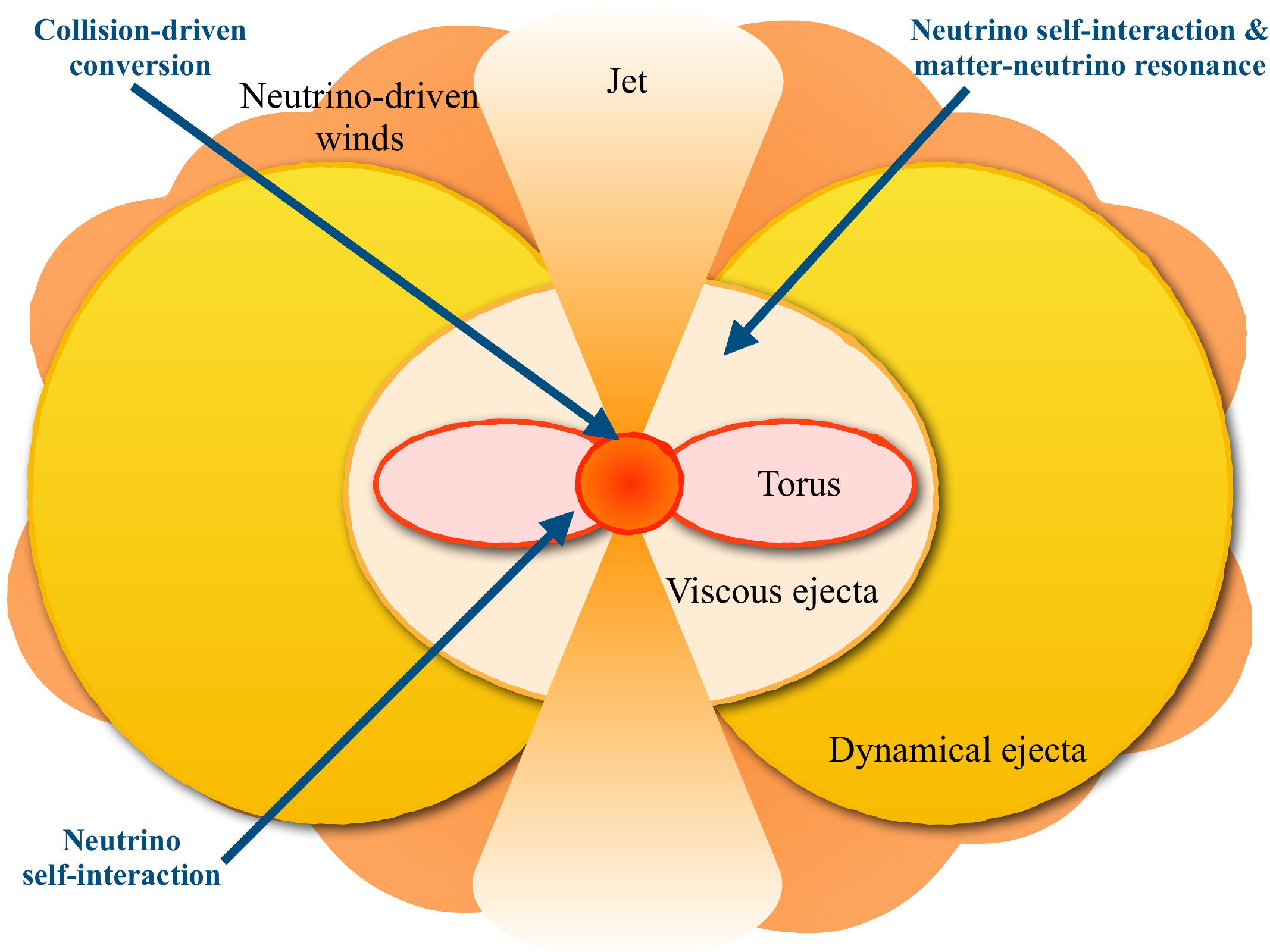}
\caption{Regions of flavor conversion of neutrinos in a core-collapse supernova (left panel) and  a neutron-star merger remnant with outflows powering a kilonova and a  gamma-ray burst (right panel).  {\it Left}: Flavor conversion enhanced by the matter background  (MSW resonances) and slow  neutrino-neutrino conversion  is expected to be relevant beyond the shock (cyan solid line). Flavor conversion mainly due to  fast instabilities and eventually collisional ones  can occur in the neutrino decoupling region (the surface of neutrino decoupling, $\nu$-sphere, is represented by the green solid line) and within the gain layer, with possible consequences on the explosion mechanism. Note that the distinction between regions of slow and fast neutrino-neutrino interactions is an oversimplification, since these two phenomena can coexist. {\it Right}: 
Flavor conversion due to neutrino-neutrino interactions and eventually collisional instabilities may affect the composition of the neutrino wind (and therefore the  nucleosynthesis) in the surroundings of the polar axis and the cooling of the disk. At larger distances from the central compact object,  because of the overall abundance of electron antineutrinos, the neutrino-neutrino term and the matter term may undergo a resonance, influencing the flavor evolution.}
\label{SN_NSM_sketch}
\end{figure}

Investigations relying on hydrodynamical simulations of the core collapse suggest that crossings in the neutrino lepton number could occur even before neutrino decoupling~\cite{Glas:2019ijo,Nagakura:2021hyb} (note, however, that the flavor solution would require a consistent evolution with the supernova hydrodynamics, and this could affect the formation of crossings).
Before neutrinos  stream freely, they may also undergo flavor conversion triggered by collisional instabilities, if the collisional rate between neutrinos and antineutrinos should be different enough~\cite{Johns:2021qby}. Potentially, flavor conversion induced by collisional instabilities could interfere with fast flavor conversion, although 
recent work suggests that, in the presence of crossings in the electron neutrino lepton number, the growth rate of the fast flavor instability is  larger than the one due to collisional instabilities~\cite{Shalgar:2023aca,Nagakura:2023xhc}. 
Yet, non-forward collisions have a non-negligible impact on  flavor conversion, enhancing or suppressing the latter according to the shape of the  angular distribution of the lepton number of neutrinos, 
or operating together with flavor conversion to shape the emerging neutrino spectra~\cite{Shalgar:2020wcx,Xiong:2024tac}.
MSW conversion is  expected to further modify the flavor distribution in the outer layers of the star, beyond the shock radius.
A visual  summary of the flavor conversion regions in the supernova interiors is displayed in the left panel of Fig.~\ref{SN_NSM_sketch}. The combination of neutrino-neutrino and MSW flavor conversions  is expected to mold the neutrino signal observed at Earth~\cite{Horiuchi:2018ofe}, as well as the nucleosynthesis in the neutrino-driven outflows~\cite{Fischer:2023ebq}. 

The  numerical solution of the seven-dimensional  neutrino quantum kinetic equations coupled to the supernova hydrodynamics is not yet available given the  technical and conceptual challenges  caused by the large difference in characteristic scales of the  terms entering the neutrino equations of motion~\cite{Tamborra:2020cul}.  
  The collapse of  massive stars is currently modeled through three-dimensional hydrodynamical simulations, where neutrino flavor conversion is neglected~\cite{Janka:2016fox,Mezzacappa:2020oyq,Burrows:2020qrp}.  Such simplification was justified in the past, also because  it was a common understanding that flavor conversion could only take place beyond the shock radius, with no implications on the explosion mechanism~\cite{Dasgupta:2011jf}.
A recent growing body of work points towards the possibility that  flavor conversion triggered by neutrino-neutrino interaction can alter the neutrino distributions of all flavors in the source core, even before neutrinos decouple from matter~\cite{Shalgar:2022rjj}. By altering the flavor ratio within the gain region, flavor conversion  could   aid or hinder the supernova explosion mechanism according to the region where flavor conversion kicks in~\cite{Ehring:2023abs,Nagakura:2023mhr}.

  \subsubsection{Detection strategies for   multi-messenger searches}
Among existing neutrino telescopes, the ones allowing for the largest supernova neutrino event rate are Super-Kamiokande~\cite{Super-Kamiokande:2024pmv} and the IceCube Neutrino Observatory~\cite{IceCube:2011cwc} (cf.~Table~1 of~\cite{Scholberg:2017czd} for an overview on  neutrino telescopes and expected event statistics). While for core-collapse supernovae in our Galaxy, IceCube would guarantee the largest statistics, Super-Kamiokande would be background free with good approximation, given the small time window characterizing the supernova neutrino burst and should allow us to reconstruct event-by-event energy information. 

Within the multi-messenger framework, the SuperNova Early Warning System (SNEWS) is an operating global network of neutrino telescopes that aims to release  a prompt alert of  core-collapse events to enable  multi-messenger observations~\cite{SNEWS:2020tbu}. In addition to being part of SNEWS, both Super-Kamiokande and IceCube are heavily involved in  the development of multi-messenger alert systems. For example, Super-Kamiokande, in collaboration with the KamLAND detector, has established an alert system that searches for  pre-supernova neutrinos and could release a warning about an upcoming electromagnetic burst about a dozen hours earlier~\cite{KamLAND:2024uia}. The IceCube supernova data acquisition system is being updated to swiftly react to alerts from the LIGO-Virgo-KAGRA gravitational wave network~\cite{Valtonen-Mattila:2023pgk}.

Hyper-Kamiokande~\cite{Hyper-Kamiokande:2018ofw} is expected to begin taking data in 2027 and the liquid scintillator Jiangmen Underground Neutrino Observatory (JUNO)~\cite{JUNO:2023dnp} should  be operative already in 2025. These neutrino telescopes
will guarantee large event statistics of supernova neutrinos with a few- to sub-degree pointing capabilities. If able to detect supernova neutrinos, the liquid-Argon Deep Underground Neutrino Experiment (DUNE)~\cite{DUNE:2020zfm}  could provide key complementarity with respect to  Hyper-Kamiokande and  JUNO, being mostly sensitive to electron neutrinos rather than electron antineutrinos. 
In addition,   experiments for the direct detection of dark matter, exploiting the coherent elastic  scattering of neutrinos on nuclei, such as XENONnT~\cite{Lang:2016zhv} at the Gran Sasso National Laboratory  and the planned neutrino detector RES-NOVA~\cite{Pattavina:2020cqc}, have the potential to complement large-scale neutrino detectors, being sensitive to all neutrino flavors. 

    \subsubsection{Supernova physics signatures imprinted in the neutrino signal}

Despite  the dependence of the neutrino signal on the progenitor mass of the collapsing star as well as the nuclear equation of state,  neutrino  detection  will be instrumental in exploring the following~\cite{Mirizzi:2015eza,Horiuchi:2018ofe}. 
\begin{itemize}
    \item {\it Alert of the stellar collapse.} Neutrinos reach Earth up to one day earlier than electromagnetic radiation.   The early detection of the supernova neutrino burst could inform astronomers of the upcoming electromagnetic burst,  enabling the  electromagnetic detection of the shock breakout~\cite{Waxman:2016qyw}. We can also use the   neutrino signal  to localize the stellar collapse in the sky 
    and  define the time window  to search for gravitational waves as well as  build a matching filter for gravitational wave searches~\cite{Nakamura:2016kkl,Drago:2023cve}.  Additionally, pre-supernova neutrinos~\cite{Patton:2017neq} produced in the  late burning stages of the stellar collapse (cf.~\cite{Farag:2020nll} for a summary of the thermal processes leading to neutrino production during the stages of stellar evolution of a massive star) could potentially provide an even earlier indication of the upcoming stellar explosion. 
    \item {\it Black hole forming collapses.} The temporal evolution of the neutrino signal carries distinctive signatures, ending abruptly at the time of black hole formation~\cite{OConnor:2010moj}. Such collapses would not be  detectable through electromagnetic radiation. Therefore, the detection of neutrinos and gravitational waves could be essential to assess the rate of black hole forming collapses, complementing ongoing electromagnetic searches~\cite{Neustadt:2021a}.
    \item {\it Pre-explosion physics.} The temporal evolution of the neutrino event rate (during the accretion phase)  carries clear imprints of the hydrodynamical instabilities characterizing the accretion phase, such as SASI and convection~\cite{Tamborra:2013laa}. Rotation can also  affect the neutrino signal~\cite{Walk:2018gaw,Takiwaki:2017tpe}.
    \item {\it Neutron star properties.} The late-time neutrino emission  (during the Kelvin-Helmoltz cooling phase)  strongly depends on the nuclear equation of state and the neutron-star properties. Despite the presence of degeneracies in the neutrino signal, its detection could place constraints on the physics of the compact remnant~\cite{GalloRosso:2018ugl,Li:2020ujl}. 
    \item {\it Physics beyond the Standard Model.} The presence of exotic signatures in the  time evolution of the neutrino signal and/or its energy distribution could be a manifestation of New Physics~\cite{Huber:2022lpm}. However, degeneracies among multiple New Physics scenarios exist  and are yet to be fully understood~\cite{MacDonald:2024vtw}.   According to the type of exotic features eventually appearing in the observed neutrino signal, before to draw any  conclusions on the discovery of New Physics relying on core-collapse supernovae,  one would  need to  model the standard core-collapse supernova scenario (including flavor conversion)  consistently--an ambitious goal yet to be achieved. 
\end{itemize}

 \subsubsection{Diffuse supernova neutrino background}
 Every second a supernova explodes somewhere in the Universe. The cumulative flux of neutrinos  from all core-collapse supernovae is named diffuse supernova neutrino background (DSNB)~\cite{Ando:2023fcc}. This  stationary and isotropic flux depends on the core-collapse supernova rate as a function of  redshift, $R(z)$, and the time-integrated (anti)neutrino  spectral energy distribution $F_\nu(E)$ emitted from each collapsing star: 
 \begin{equation}
 {\Phi_{\nu}}(E) = \int_0^\infty dz\  F_{\nu}(E_z)  \frac{R(z)}{H(z)}\ ,
 \end{equation}
with  $E_z = (1+z) E$ being the  redshifted neutrino energy, and $H(z)$ is the Hubble parameter.

To date, the largest uncertainty plaguing the forecast of the DSNB signal  is the core-collapse supernova rate~\cite{Ekanger:2023qzw}, while the DSNB negligibly depends on the uncertainties on the initial mass function~\cite{Ziegler:2022ivq}.   The unknown rate of collapses leading to  black hole formation~\cite{Lunardini:2009ya} and magnetorotational collapses~\cite{Martinez-Mirave:2024zck}, as well as the metallicity evolution of the host  galaxies~\cite{Ashida:2023heb}, and the impact of mass transfer during binary evolution   (if the collapsing star happens to be in a binary system)~\cite{Horiuchi:2020jnc} are some of the uncertainties in  the forecast of the DSNB. In addition, the unknowns linked to the  supernova microphysics (such as the neutron star mass, high-density nuclear equation of state, and 
 the flavor conversion physics~\cite{Kresse:2020nto,Lunardini:2012ne}) also affect the DSNB.

The upcoming DSNB detection  will be instrumental in placing neutrino-based constraints on the properties of the supernova population, the average redshift distribution of supernovae in the local Universe, and the average supernova energetics. In particular, the rate of core-collapse supernovae measured through electromagnetic surveys could be biased because electromagnetically dark  collapses could be missed, but such collapses would contribute to the DNSB~\cite{Lien:2010yb}.

The  Gadolinium loading of the water Cherenkov detector Super-Kamiokande  has drastically enhanced the signal over background ratio, already excluding some of the most optimistic DSNB models  as well as hinting towards a signal excess at $2.3 \sigma$~\cite{Super-Kamiokande:2023xup,SK-Gd:2024}. The latest Super-Kamiokande upper limits on  $\bar\nu_e$'s of astrophysical origin are presented in the right panel of Fig.~\ref{fluence_DSNB} together with the expected DSNB signal.
In the  future, Hyper-Kamiokande~\cite{Hyper-Kamiokande:2018ofw}, possibly loaded with Gadolinium, and JUNO~\cite{JUNO:2022lpc}   will  allow us to investigate  the DSNB physics with larger statistics (cf.~Table~2 of~\cite{Martinez-Mirave:2024zck} for estimations of the expected neutrino number of events). On the other hand, neutrino  and direct-detection dark matter experiments, exploiting the coherent elastic neutrino scattering on nuclei, can  place upper limits on the DSNB emission in the non-electron flavors~\cite{Suliga:2021hek}.

\subsection{Neutron-star merger remnants}\label{sec:NSM}

\subsubsection{Neutron-star merger physics}

 In 2017, the joint detection of gravitational waves and electromagnetic radiation from the  merger of two neutron stars  confirmed the relevance of these objects for astrophysics and fundamental physics~\cite{Margutti:2020xbo,Metzger:2019zeh}; see Table~\ref{tab:summary} for a summary of the multi-messenger energetics. The detection in 2017 has provided 
 a strong indication that at least a fraction of the elements heavier than iron can be synthesized in these sources. Yet, several aspects of the physics of merger remnants remain to be clarified, such as the interplay between magnetic fields and neutrino transport, and their degree of sophistication~\cite{Foucart:2024npn,Kiuchi:2023obe}.

Following the merger event, the accretion disk undergoes cooling via neutrino emission, with neutrino-matter interactions determining the composition of the merger outflows~\cite{Janka:2022krt}. According to the mass of the system and the nuclear equation of state, the central compact object (a newly-born neutron star) can immediately collapse 
 into a black hole or it can be a rapidly spinning neutron star remnant. In the latter scenario,  a hyper-massive~\cite{Baumgarte:1999cq} or  supra-massive~\cite{1992ApJ...398..203C} neutron star may form, which is expected to collapse into a 
black hole in  $\gtrsim 1$~s  from the merger,  
it is also foreseen that  a stable neutron star may be born~\cite{Shibata:2019wef}.

The ejecta emitted in connection with the merger event have different origin and composition~\cite{Janka:2022krt,Shibata:2019wef}, as displayed in the right panel of Fig.~\ref{SN_NSM_sketch}. During the merger,  dynamical ejecta form. These  consist of shock-driven ejecta (produced during the collision of the two neutron stars, with outflows expanding across all directions and especially in the polar region) and tidal ejecta (due to mass outflows confined in the proximity of the equatorial plane and gaining sufficient angular momentum to be ejected via  pre-merger tidal forces), with the tidal component being characterized by a lower electron fraction since its matter is not affected by shock heating and neutrino irradiation. More massive ejecta can originate from the unbinding of the outer layers of the newly born stable or metastable neutron star caused by the build-up of magnetic pressure.  

The accretion disk surrounding the central compact object can also produce an outflow, whose properties depend on the nature of the compact object, and is mainly driven by viscous heating and nuclear energy release~\cite{Fernandez:2013tya}. 
If the central compact object is a black hole, then the disk is  dense enough to be initially opaque for neutrinos. As the temperature  decreases, neutrino cooling tends to become inefficient, and viscous heating in the accretion disk leads to the isotropic ejection of material. If the central compact object is a neutron star, then a neutrino-driven wind expands in the proximity of the polar region, in addition to the viscous-driven disk ejecta.

The observation of the gamma-ray burst GRB 170817A in connection with the gravitational wave event GW170817~\cite{LIGOScientific:2017zic} has provided the first observational proof that the spinning black hole at the center of the merger event can  harbor a Poynting-flux dominated outflow observed as short gamma-ray burst~\cite{Eichler:1989ve}. 
 Long and short gamma-ray bursts have been  assumed to originate from collapsing massive stars and neutron-star mergers, respectively. However, recent observations challenge this paradigm~\cite{2024Natur.626..742Y,Rastinejad:2022zbg}, with novel ideas being proposed  to explain the association of kilonovae to long gamma-ray bursts~\cite{Gottlieb:2023sja,Cheong:2024hrd}.

 \subsubsection{Impact of neutrino physics on multi-messenger observables}

 Neutrinos are fundamental  in the very hot and dense regions formed following the collision of two neutron stars~\cite{Foucart:2024cjr}, with neutrino physics affecting the thermodynamics of the remnant as well as the composition of the outflows. The neutrino density in the core of neutron-star merger remnants is roughly comparable to the one characterizing the supernova core, although rapidly dropping in time and with an overall abundance of electron antineutrinos over neutrinos (due to the neutron richness of the environment). We expect neutrinos and antineutrinos, mostly of electron flavors, to be copiously emitted up to tens of seconds from the coalescence of two neutron stars, with an average energy of $\mathcal{O}(10)$~MeV (cf.~Table~\ref{tab:summary}). 
 
 The detection chances of thermal neutrinos from neutron-star merger remnants   are poor~\cite{Kyutoku:2017wnb}; in fact, assuming that  the ratio between neutron-star mergers and core-collapse supernovae  in our Galaxy is comparable to the one obtained from the local rates of such sources--see Table~\ref{tab:summary}--we expect a neutron-star merger to be $10^{-4}$--$10^{-2}$ less frequent than a  supernova. This is  also visible from the left panel of Fig.~\ref{fluence_DSNB} where the orange lines, representing the fluence from a Galactic neutron-star merger remnant (note that the distance of $10$~kpc has been chosen for representative purposes), is contrasted with  the one from a  core-collapse supernova at the same distance from Earth. 
 
 The modeling of neutrino flavor conversion in the core of neutron-star merger remnants is as challenging as for core-collapse supernovae, with additional complications linked to the disk geometry~\cite{Tamborra:2020cul}. Similar to the supernova case, fast and collisional instabilities can take place  in the proximity of the neutrino decoupling region, as shown in the right panel of Fig.~\ref{SN_NSM_sketch}, triggering flavor conversion. In addition, because of the abundance of electron antineutrinos over neutrinos, the neutrino-neutrino self-interaction term can have opposite sign with respect to the matter term in the Hamiltonian; the cancellation between these two potentials can  lead to a flavor conversion phenomenon known as {\it matter-neutrino resonance} at larger distances from the central compact object~\cite{Malkus:2012ts}. However, recent work highlights the importance of the modeling of the neutrino angular distributions  to reliably assess whether any flavor conversion should be expected from the matter-neutrino resonance~\cite{Padilla-Gay:2024wyo}.   
 
 While general relativistic neutrino magnetohydrodynamic simulations of neutron-star merger remnants do not take into account neutrino flavor conversion, the latter likely  has   implications on the synthesis of the  elements heavier than iron via the rapid neutron capture process ($r$-process). Hence, neutrino flavor conversion can affect the electromagnetic kilonova properties, modifying the production rate of lanthanides especially in the proximity of the polar region~\cite{Wu:2017drk,Just:2022flt,Fernandez:2022yyv,Li:2021vqj}. 

The multi-messenger data from the GW 170817 event can already be employed to place constraints on the  existence of New Physics, also linked to the neutrino sector. However, as it is the case for core-collapse supernovae, it is crucial to first gain control on the standard physics involved in neutron-star mergers, especially for what concerns neutrino transport and the physics linked to the magnetohydrodynamic instabilities, which still involves several approximations~\cite{Foucart:2024npn,Ciolfi:2020cpf,Hayashi:2022cdq}.

\subsubsection{Diffuse neutrino background from neutron-star mergers}

To simulate the  binary merger population diversity, we should  rely on a large suite of long-term general-relativistic neutrino-radiation magnetohydrodynamics merger simulations. These are not yet available.  Nevertheless, taking into account the very large uncertainties on the cosmic rate of neutron-star mergers~\cite{Mandel:2021smh} (which predict  a rate of neutron-star mergers that is one-to-two orders of magnitude smaller than the supernova one, or even smaller, cf.~Table~\ref{tab:summary}) and on the neutrino fluence from each merger event (which is expected to be comparable to or smaller than the core-collapse supernova one, see the left panel of Fig.~\ref{fluence_DSNB}), the diffuse neutrino flux from these sources should be at best one-to-three orders of magnitude smaller than the DSNB one (cf.~the right panel of Fig.~\ref{fluence_DSNB}), as already suggested by preliminary estimations~\cite{Schilbach:2018bsg}.   Hence, thermal neutrinos from neutron-star merger remnants 
cannot be adopted to probe  the  merger remnant population.

 \begin{bclogo}[
  couleur=bgblue,
  arrondi=0,
   logo= ,
  barre=none,
  noborder=true]{Neutrinos from the  interiors of explosive transients: key points}
  \indent
  \begin{itemize}
  \item[ ] $\ast$ Neutrinos are fundamental to the core collapse of a massive star and carry $99\%$ of the supernova binding energy.
  \item[ ] $\ast$ Neutrino flavor conversion is expected to take place in the core of a supernova, potentially affecting the explosion mechanism and the related multi-messenger emission. Yet, neutrino flavor conversion is not  taken into account in hydrodynamic simulations. 
  \item[ ] $\ast$ The detection of neutrinos from a Galactic supernova will be essential to alert observers focusing on the electromagnetic spectrum about the upcoming collapse,  provide information on the pre-explosion physics (complementing the input coming from gravitational waves), as well as the nature and properties of the central compact object. 
  \item[ ] $\ast$ The detection of the diffuse emission of neutrinos from all supernovae in our Universe (the DSNB) would be key to gain insight  on the properties of the population of collapsing massive stars, complementing electromagnetic data. 
 \item[ ] $\ast$ Neutrinos are as abundant in core-collapse supernovae as in neutron-star merger remnants, with electron neutrinos having a larger local number density than electron neutrinos in  merger remnants. Neutrinos and their flavor conversion physics can influence the abundance/species of nuclei synthetized via the $r$-process.
\item[ ] $\ast$ The detection of  neutrinos from these neutron-star merger remnants is unlikely because of the low rate of such events expected in our Galaxy. The diffuse flux of   neutrinos from neutron-star mergers is negligible with respect to the DSNB.
  \end{itemize}
\end{bclogo}

\section{Neutrinos from the outskirts  of explosive transients}\label{sec:outskirt}
 
In this section, the non-thermal electromagnetic and neutrino emission stemming in the aftermath of supernovae as well as 
neutron-star mergers is discussed, including newly discovered transients that may originate from these source classes. In particular, we highlight the opportunities to infer the source physics by relying on the multi-messenger detections  as well as optimal detection strategies.

\subsection{Non-thermal neutrino emission from transient astrophysical sources}

The IceCube Neutrino Observatory has detected  neutrinos of astrophysical origin with $\rm{TeV}$--$\rm{PeV}$ energy since  2013. 
A handful of statistically significant neutrino-electromagnetic associations has been found, e.g.~in coincidence with the blazar event TXS 0506$+$056, the active galaxy NGC 1068, and our own Galaxy~\cite{IceCube:2018dnn,IceCube:2022der,IceCube:2023ame}.   The potential for multi-messenger follow-up in real-time using neutrino alert is steadily improving~\cite{IceCube:2023agq}; at the same time, the overall  number of detected high-energy neutrino events increases~\cite{IceCube:2025zyb}, offering additional opportunities to search for    multi-messenger associations.
A very intriguing opportunity would stem from the detection of high-energy neutrinos from transients originating from collapsing massive stars and neutron-star mergers. In fact, such neutrinos can shed light on the mechanism powering the observed electromagnetic emission as well as the properties of the  source and its surrounding medium~\cite{Guarini:2023rnd,Fang:2020bkm}.

\subsubsection{Collapsing massive stars and emerging transient sources}
\begin{figure}[t]
\centering
\includegraphics[width=.45 \linewidth]{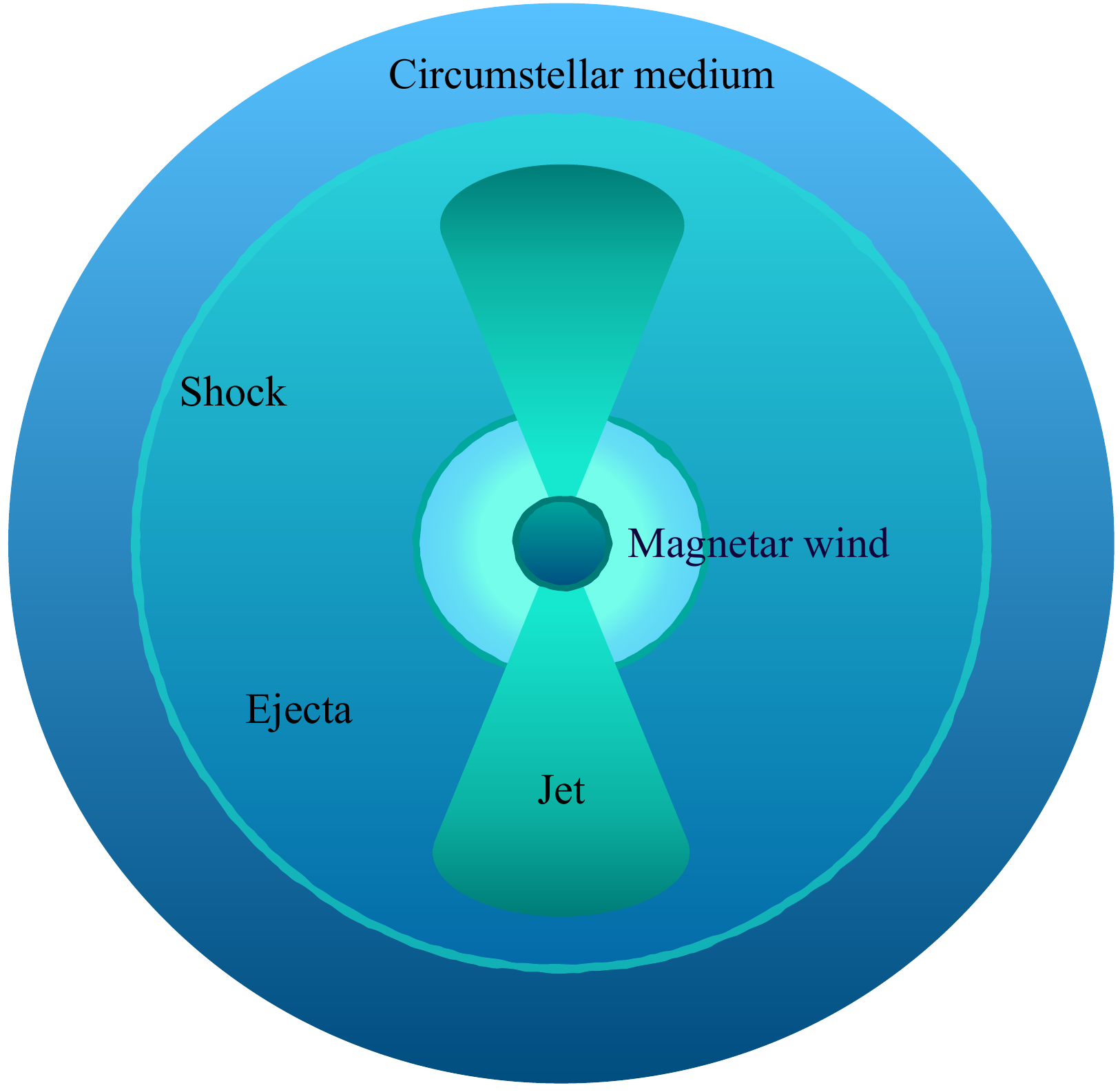}
\includegraphics[width=.45 \linewidth]{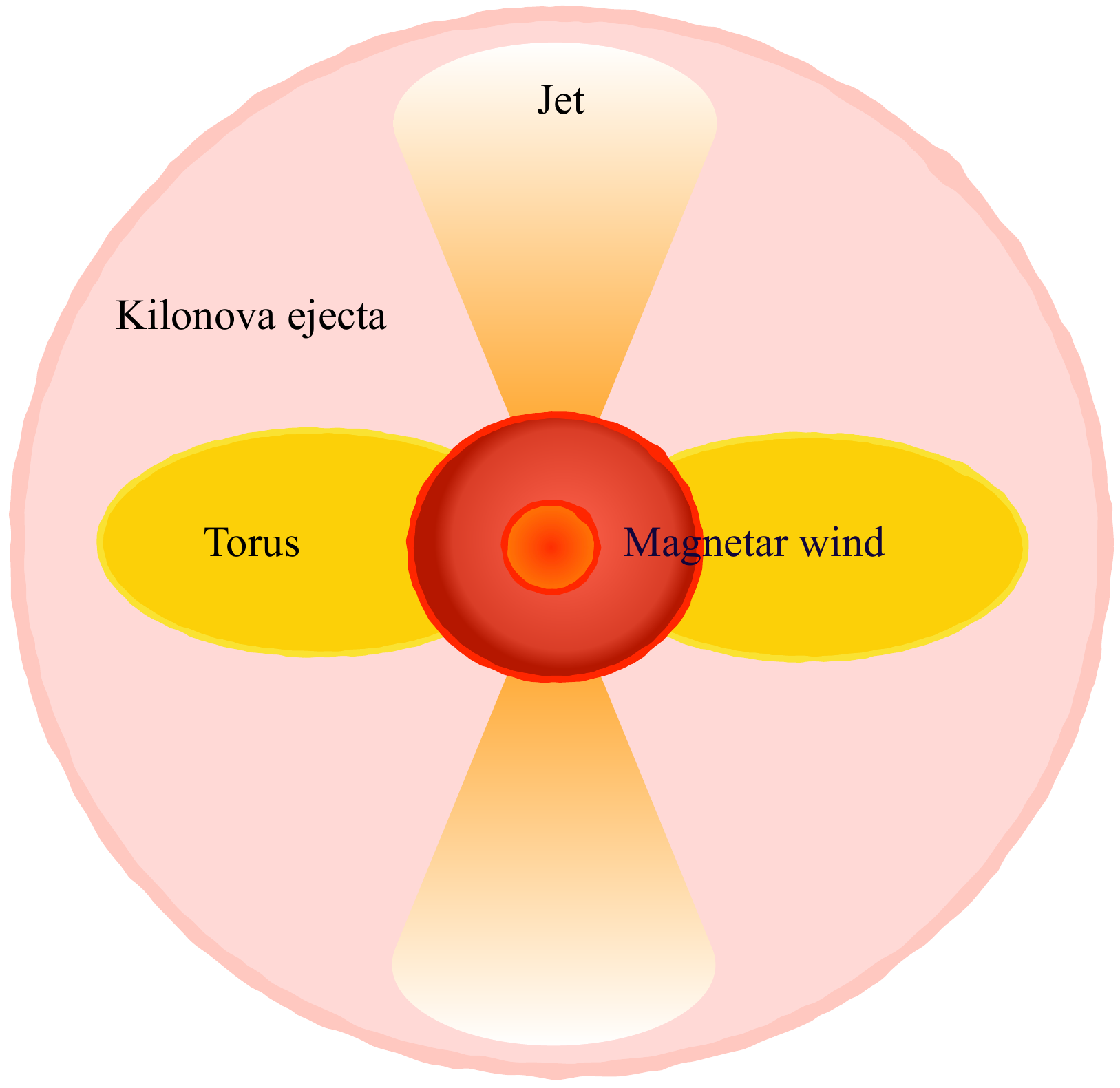}
\caption{Sketch of the sites of non-thermal neutrino production in a collapsing massive star (left panel) and a neutron-star merger remnant (right panel). {\it Left}: Outflows launched during the core collapse and powered by a central source of heating, such as a magnetar  (emerald green region).   A jet can be harbored and eventually  break out (aqua region), being observable as a gamma-ray burst. The supernova ejecta (light blue region) can interact with the dense circumstellar medium (in blue) as the   shock  breaks out from the circumstellar medium. Particle acceleration, and therefore  neutrino production, can occur  along the jet, in the magnetar wind, and as the external shock propagates in the circumstellar medium. 
  {\it Right}: Outflows launched in the aftermath of a neutron-star merger, the central power engine  could be a magnetar (orange region). A jet (in light orange) can  break out leading to a  gamma-ray burst. Efficient neutrino production, as a result of particle acceleration, can take place along the jet, in the magnetar wind (red region), as well as because of  the interaction between the  jet and the ejecta (pink region). 
\label{SN_NSM_sketch_HE}}
\end{figure}

In addition to thermal neutrinos with energy of $\mathcal{O}(10)$~MeV, the explosion resulting from the collapse of a massive star can also lead to the production of non-thermal neutrinos  with higher energy (in the TeV--PeV range). Such neutrinos are produced because of the interactions of accelerated protons with other protons or photons (hadronic or photo-hadronic interactions)~\cite{Kelner:2006tc,Kelner:2008ke}.
A summary of the  particle acceleration sites, leading to non-thermal  neutrino production for high-energy transients stemming from collapsing massive stars, is provided in the left panel of Fig.~\ref{SN_NSM_sketch_HE}.

\begin{figure}[t]
\centering
\includegraphics[width=.6 \linewidth]{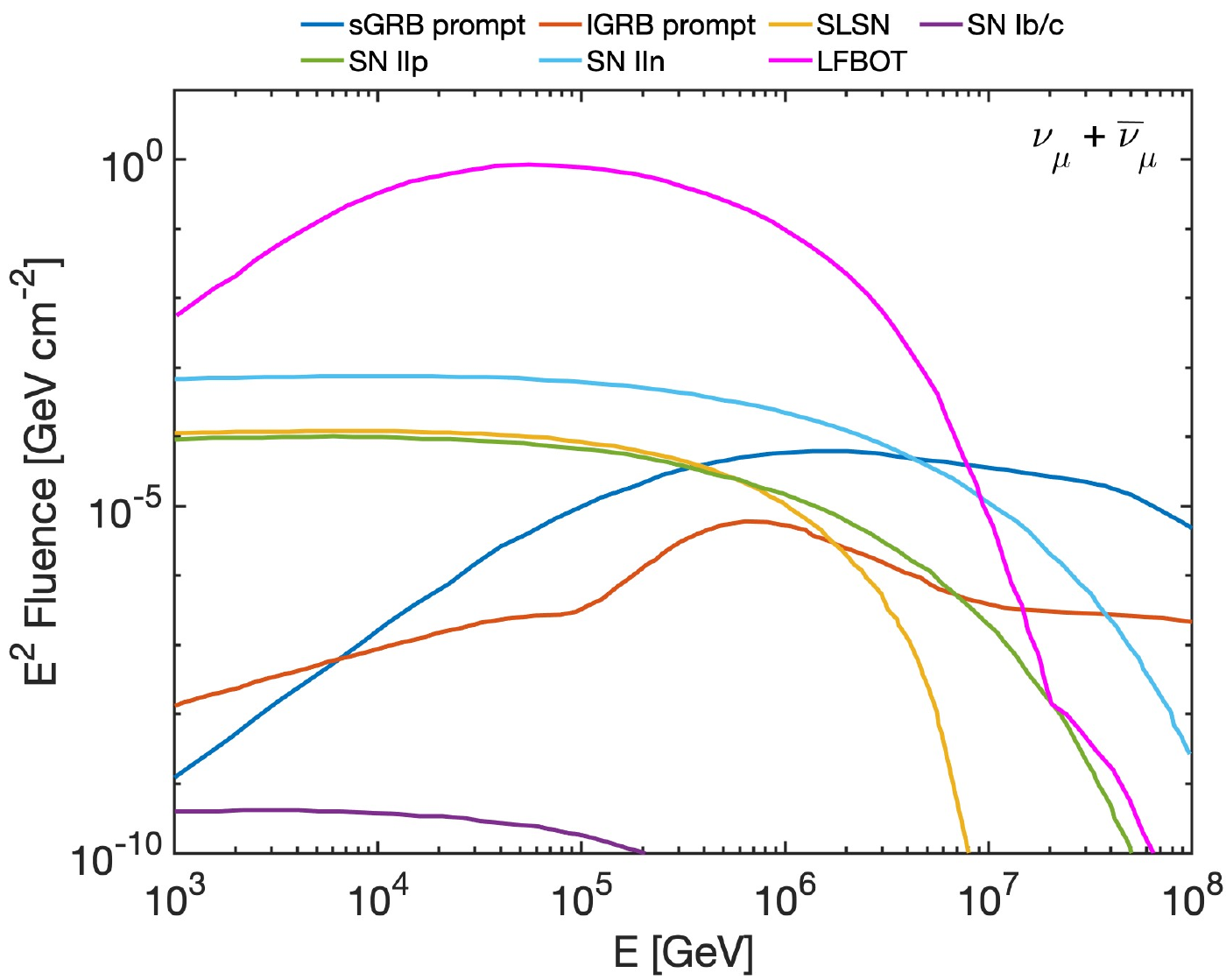}
\caption{Muon neutrino and antineutrino fluences  as functions of the neutrino energy expected for representative transients stemming from collapsing massive stars and neutron-star mergers (located ad different distances from Earth). 
The following curves are plotted: prompt emission from a short gamma-ray burst  (sGRB, GRB 170817A, $z= 0.008$)~\cite{Biehl:2017qen}, the prompt emission from a long gamma-ray burst (lGRB, $z=2$, for a magnetized jet with gradual dissipation)~\cite{Pitik:2021xhb}, a superluminous supernova (SLSN AT2019fdr, $z= 0.27$)~\cite{Pitik:2021dyf},  supernovae of Type Ib/c, IIp and IIn ($z=0.002$)~\cite{Sarmah:2022vra}, as well as  a luminous fast blue optical transient (LFBOT AT2018cow, $z = 0.014$)~\cite{Guarini:2022uyp}. These curves should be considered for orientation since uncertainties on the astrophysical parameters and plasma properties can lead to changes by a few orders of magnitude in the neutrino fluence.
\label{HEfluence} }
\end{figure}

The supernova ejecta expand at $3$--$10 \times 10^3$~km/s, and diffusive shock acceleration can take place in the proximity of the  shock. In   shock-powered supernovae, high-energy neutrinos could be produced  because of the {\it interaction of  shock-accelerated protons with  protons from the dense circumstellar medium} (gamma-ray emission should be expected as well), or with the photon spectrum at the external shock (albeit the related neutrino contribution is subleading for non-relativistic or mildly-relativistic shocks)--see the left panel of Fig.~\ref{SN_NSM_sketch_HE}. Such non-thermal particle emission is  efficient  for supernovae of Type IIn, IIp, or superluminous supernovae (cf.~Table~\ref{tab:summary} and Fig.~\ref{HEfluence}), whose circumstellar medium is expected to be especially dense because of mass-loss ejection from the progenitor star~\cite{Murase:2010cu,Sarmah:2022vra,Pitik:2023vcg}.   
High-energy neutrinos from these sources have not  been detected yet, with  the IceCube Neutrino Observatory  providing upper limits on the diffuse emission from (shock-powered) supernovae~\cite{IceCube:2023esf}. However, the detection of  these neutrinos would be important to  explore   the structure
and properties of the circumstellar medium as well as the ejecta.
Interacting supernovae can  be among the main contributors to the  diffuse neutrino flux between $10$ and $10^3$~TeV, without overshooting  the isotropic gamma-ray background observed by Fermi-LAT~\cite{Sarmah:2022vra,Waxman:2024njn}.

If the central compact object is a  magnetar, we   expect neutrino emission linked to its spin down~\cite{Fang:2018hjp}. 
In this case, 
the protons accelerated in the {\it magnetar wind}  can undergo hadronic or photo-hadronic interactions with the thermal optical photons and the non-thermal X-rays produced in the    ejecta forming a  wind nebula, as shown in Fig.~\ref{SN_NSM_sketch_HE}.

A {\it jet} can be harbored in  stripped supernovae (of Type Ib/c), leading to  gamma-ray bursts (cf.~Table~\ref{tab:summary} and Fig.~\ref{SN_NSM_sketch_HE}). The jet may be successful or not, according to its energy and the density of the stellar envelope.
Below the photosphere, the mixing between the jet and the cocoon loads the jet with baryons, inhibiting  subphotospheric collisionless shocks~\cite{Gottlieb:2021pzr}. However, if the jet is even mildly magnetized, magnetic reconnection and subshocks may favor   particle acceleration and therefore neutrino production in the GeV--TeV range~\cite{Guarini:2022hry,Rudolph:2023auv}. 
  In the optically thin region of the relativistic jet, observed as a gamma-ray burst, the accelerated protons can 
interact with  non-thermal photons to produce high-energy neutrinos~\cite{Pitik:2021xhb,Kimura:2022zyg}, as displayed in Fig.~\ref{HEfluence}. 
Notably the efficiency of particle acceleration depends on  whether heavier nuclei can be entrained in the jet, with neutrino emission  becoming less efficient~\cite{Horiuchi:2012by,DeLia:2024kjv,Biehl:2017zlw}. 
 To date, neutrinos emitted from gamma-ray bursts have not yet been detected, these findings are however compatible with theoretical estimations~\cite{IceCube:2023woj, Pitik:2021xhb}.

The prospects of detecting neutrinos from newly discovered source classes potentially linked to collapsing massive stars are also very exciting.
An example is provided by luminous fast blue optical transients (LFBOTs, see Table~\ref{tab:summary}), such as AT2018cow~\cite{Ho:2023sfv}. The origin of these transients is not yet known, but they are expected to be powered by a compact object, launching an asymmetric and fast outflow~\cite{Cao:2024yao}. It has been proposed that luminous fast blue optical transients  may stem from collapsing massive stars stripped of their hydrogen envelope and harboring a choked jet surrounded by an inflated cocoon~\cite{Gottlieb:2022old}; the related  neutrino fluence is plotted in Fig.~\ref{HEfluence} for reference.  Alternatively, the luminous fast blue optical transient electromagnetic emission could come from the delayed merger of a black hole with a Wolf-Rayet star~\cite{Metzger:2022xep}. Neutrinos could be produced because of the interaction between the outflow and the circumstellar medium~\cite{Guarini:2022uyp} or in the magnetar wind~\cite{Fang:2018hjp}. The detection of  neutrinos  would be crucial to breaking the degeneracies plaguing the electromagnetic data (which currently do not allow for a robust assessment of the source properties), allowing to pinpoint the mechanism powering the observed electromagnetic emission~\cite{Guarini:2022uyp}.

\subsubsection{Neutron-star mergers}

Following up the merger of two neutron stars, if the remnant compact object is a  {\it magnetar}, its wind can interact with the ejecta, with protons being accelerated at the magnetosphere or the termination shock of the wind, see the right panel of Fig.~\ref{SN_NSM_sketch_HE}. While particle acceleration is not efficient at the termination shock, neutrino emission from the magnetar  could be detected in connection with gravitational waves~\cite{Fang:2017tla}; however, recent work suggests that the detection of such neutrinos may be more challenging than initially expected~\cite{Mukhopadhyay:2024ehs}.

The non-detection of neutrinos from  GRB 170817A~\cite{ANTARES:2017bia}, observed in connection with gravitational waves,  is consistent with the fact that the {\it relativistic jet} was off-axis with respect to the observer~\cite{Gottlieb:2021pzr,Biehl:2017qen} (cf.~Fig.~\ref{HEfluence} for an estimation of the muon neutrino fluence expected from the prompt emission). As the jet propagates within the ejecta, particle acceleration can take place at  internal shocks inside the ejecta, leading to a neutrino signal within the detection reach of  the IceCube Neutrino Telescope~\cite{Kimura:2018vvz}. Table~\ref{tab:summary} provides a summary of the energetics and the rate of the transients linked to the coalescence of neutron stars, with the related acceleration sites being summarized in the right panel of Fig.~\ref{SN_NSM_sketch_HE}.

\subsection{Best detection strategies for upcoming high-energy multi-messenger searches}

The detection (or lack thereof) of high-energy neutrinos from transients stemming from collapsing massive stars or neutron-star mergers allows to gain crucial insight about the source properties, the behavior of plasma in highly magnetized media, as well as the efficiency of particle acceleration. Of course, the information provided by high-energy neutrinos should be complemented by the one carried by photons across wavebands and gravitational waves. Notably, detecting non-thermal neutrinos is especially crucial for  electromagnetically dark sources, such as choked jets. In this case, we may expect a correlation of high-energy neutrinos with gravitational waves and  $\mathcal{O}(10)$~MeV neutrinos (or interactions among such neutrinos~\cite{Guo:2023sbt}) produced  in the source core.

Within the next decade, the number of electromagnetic detections of transients is expected to grow swiftly thanks to wide-field and high-cadence surveys, such as  the soon-to-be operative Vera C.~Rubin Observatory~\cite{LSST:2022kad}  and ULTRASAT~\cite{Shvartzvald:2023ofi}, the Nancy Grace Roman Space Telescope~\cite{WFIRST:2018mpe} as well as the Cherenkov Telescope Array (CTA)~\cite{CTAConsortium:2010umy}. These surveys,  among others, will complement the pioneering findings of  the  Zwicky Transient Facility (ZTF)~\cite{Dekany:2020tyb}, the Panoramic Survey Telescope and Rapid Response System 1 (Pan-STARRS1)~\cite{Chambers:2016jzn},  the Swift-Ultra-Violet/Optical Telescope~\cite{Roming:2005hv},  and the  James Webb Space Telescope~\cite{JWST:2023jqa}. 
At the same time, the next-generation gravitational-wave detectors, e.g.~the Cosmic Explorer~\cite{Evans:2023euw} and Einstein Telescope~\cite{Badaracco:2024kpm}, are expected to  largely expand the observational horizon explored by  LIGO, Virgo, and KAGRA~\cite{KAGRA:2013rdx}.
For what concerns the  detection of high-energy neutrino counterparts, the current observational landscape is dominated by the IceCube Neutrino Telescope~\cite{Kurahashi:2022utm} and  the under-construction Cubic Kilometre Neutrino Telescope (KM3NeT)~\cite{KM3NeT:2024paj}. The planned IceCube-Gen2~\cite{IceCube-Gen2:2020qha}, the Giant Radio Array for Neutrino Detection (GRAND200k)~\cite{GRAND:2018iaj}, and the Pacific Ocean Neutrino Experiment (P-ONE)~\cite{P-ONE:2020ljt}, among several other neutrino detectors, are expected to strengthen and  expand our detection opportunities (cf.~Table 1 of~\cite{Guepin:2022qpl} for an overview of existing and planned high-energy neutrino telescopes). 

To investigate  the source properties and test our understanding of particle acceleration physics, it is however necessary to  facilitate coordination as well as data exchange through multi-messenger programs. Several joint  programs are already operative, for example ZTF, Pan-STARRS1, the Dark Energy Survey (DES), and  the All-Sky Automated Survey for SuperNovae (ASAS-SN) carry out target-of-opportunity searches for optical counterparts of neutrino events and  {\color{red}vice versa~\cite{IceCube:2020mzw,Stein:2022rvc,Pan-STARRS:2019szg,ASAS-SN:2022gst,DES:2019hqa};} similar collaborations exist for neutrino and gamma-ray joint searches, involving Fermi-LAT  and the Imaging Atmospheric Cherenkov Telescopes (IACTs)~\cite{VERITAS:2021mjg,Garrappa:2021ihz}.
To  take full advantage of upcoming joint high-energy neutrino and electromagnetic  detections, it is crucial to strengthen correlations between neutrinos, the radio waveband (likely carrying imprints of  interactions with the circumstellar medium), and the X-ray one (sensitive to the activity of the central engine)   to break the degeneracies intrinsic  to the optical-ultraviolet-infrared band~\cite{Pitik:2023vcg,Guarini:2023rnd,Waxman:2024njn}. 
The time window adopted to look for neutrinos should not be  chosen arbitrarily; in fact, we expect that the neutrino signal  can only  be tightly linked to the waveband whose photons are produced in the same source region.

 \begin{bclogo}[
  couleur=bgblue,
  arrondi=0,
   logo= ,
  barre=none,
  noborder=true]{Neutrinos from the stellar outskirts: key points}
  \indent
  \begin{itemize}
  \item[ ] $\ast$ Neutrinos can be produced in high-energy astrophysical transients thanks to the interaction of accelerated protons with the photon and/or baryon backgrounds.
  \item[ ] $\ast$ In the aftermath of the core collapse of a massive star or the merger of two neutron stars, non-thermal neutrinos could be produced along the jet, in the magnetar wind, as the shock propagates in the circumstellar medium, and possibly thanks to the interaction of the ejecta with the jet.
  \item[ ] $\ast$ The non-thermal neutrino signal carries  signatures of the source power engine. Therefore it allows to probe the source physics and plasma physics in magnetized media. 
   \item[ ] $\ast$ Multi-messenger transient astronomy is undergoing a revolution with an  unprecedented amount
of data.  It is necessary to swiftly develop the infrastructure necessary to  cross-correlate  and analyze large multi-messenger data sets. 
  \end{itemize}
\end{bclogo}

\section{Neutrinos and explosive transients: future challenges and opportunities}
\label{sec:conclusions}

Despite their weakly interacting nature, neutrinos are crucial particles in the core-collapse explosion mechanism. Because of their impact on the electron fraction, neutrinos also contribute to regulating the production of isotopes heavier than iron in neutron-star merger remnants as well as core-collapse supernovae. In combination with gravitational waves, thermal neutrinos are  crucial probes of the physics governing the innermost regions of collapsing massive stars, providing unique information on the (pre-)explosion physics. The non-thermal emission of electromagnetic radiation and  neutrinos from the source outskirts, instead, could provide crucial insight into the structure of the circumstellar medium as well as the plasma properties, complementing  the information carried by the electromagnetic signal.

The growing number of likely multi-messenger detections of transient astrophysical sources, as well as the major developments taking place  in the field of neutrino telescopes and  gravitational wave detectors, jointly with the upcoming  wide-field and high-cadence surveys revolutionizing electromagnetic observations of transients,
call for an improved understanding of the source physics and  related particle production. Only by advancing the theoretical modeling, we would be able to take advantage of the upcoming plethora of data.
On the one hand, it is essential to make progress in the modeling of neutrino flavor conversion in dense media. One challenge is that  neutrino flavor conversion is not taken into account in (magneto)hydrodynamic simulations of the source. Hence, the impact of the feedback of  flavor conversion physics on the source and  the nucleosynthesis is  yet to be understood. On the other hand, it is  crucial to swiftly gain ground in our understanding of particle acceleration. Recent work  highlights the importance of coupling  particle transport consistently with general relativistic magnetohydrodynamics and particle-in-cell simulations  to reliably assess the regions (and the  efficiency) of particle acceleration and 
forecast the multi-messenger signals. 
For emerging source classes, 
 neutrinos  will be important to  breaking the degeneracies affecting  the electromagnetic signal, pinpointing the physics mechanisms powering such sources. 

On the observational side, the next decade will see a revolution in transient astronomy with an unprecedented amount of data. In order to take full advantage of such opportunities  
and guide neutrino searches, electromagnetic observations across different wavelengths,  especially in the radio and X-ray wavebands, should be strengthened.  
In addition, the temporal information encoded in the thermal neutrino signal could be used to facilitate gravitational wave searches as well as to extract information about the physics of the source core which would complement the gravitational wave one.

The unprecedented number of real-time alerts expected to become available to telescopes involved in multi-messenger searches calls for a rapid development of the cyberinfrastructure needed to coordinate, cross-correlate, and analyze the large amount of data provided by such telescopes. In this respect,  optimization of the  communication infrastructure adopted to broadcast  alerts is urgently needed. The Scalable Cyberinfrastructure to support Multi-Messenger Astrophysics (SCiMMA)~\cite{SCiMMA:2021kab} and the Time-domain Astronomy Coordination Hub (TACH)~\cite{2020AAS...23510715S} are examples of projects in this direction.

Expanding the detection horizon beyond a single source,  the diffuse emission of neutrinos from specific source classes can inform us on the source population properties. Besides complementing electromagnetic measurements, diffuse neutrino data could provide  input on the sources that are dim or dark electromagnetically.  However, to be able to extract details from the neutrino signal and predict the diffuse signal reliably, we should proceed  on the modeling of stellar evolution, as well as  better understand the role of a  binary companion in the stellar evolution, and the cosmic rate.

\section*{Acknowledgements}
I am especially grateful to  Pablo Mart\'{i}nez-Mirav{\'e}, Tetyana Pitik, Georg Raffelt, and Meng-Ru Wu for   insightful feedback on the manuscript, as well as Kenta Kiuchi and Daniel Kresse for helpful discussions. 
The literature on this subject  evolves at a very fast pace, I apologize in advance for any contribution that I could not  adequately cover in this review. 
Support from the Danmarks Frie Forskningsfond (Project No.~8049-00038B), the Carlsberg Foundation (CF18-0183), the Villum Foundation (Project No.~13164),  the European Union (ERC, ANET, Project No.~101087058), and the Deutsche Forschungsgemeinschaft through Sonderforschungsbereich SFB 1258 ``Neutrinos and Dark Matter in Astro- and Particle Physics'' (NDM) is acknowledged.

\section*{Competing interests}
The author declares no competing interests. 

\section*{Publisher’s note}
Springer Nature remains neutral with regard to jurisdictional claims in published maps and institutional affiliations.

\bibliography{biblio}

\end{document}